# Development of a Multi-Agent System for Optimal Sizing of a Commercial Complex Microgrid

Soheil Mohseni
Department of Electrical Engineering, University Campus 2, University of Guilan
Rasht, Iran
soheilmohseni92@gmail.com

Seyed Masoud Moghaddas Tafreshi
Department of Electrical Engineering, Faculty of Engineering, University of Guilan
Rasht, Iran
tafreshi@guilan.ac.ir

*Abstract*—In this paper, a novel intelligent method based on a multi-agent system (MAS) is applied to the problem of optimal sizing in a stand-alone office complex microgrid such that the electricity demand of the office building and the charging demand of the plug-in hybrid electric vehicle (PHEV) charging station are met. The proposed MAS-based architecture consists of five different agents, namely generation agent (GA), electrical load agent (LA), PHEV charging station agent (SA), control agent (CA), and design agent (DA) which are organized in three levels. In the proposed MAS, control agent coordinates the interactions between the generation agent and electrical load and charging station agents and following a request from design agent, sends the information on the operation of the microgrid with determined sizes to it. According to the received information, design agent finds the optimal sizes of the system's components such that the electricity and charging demands are met considering two reliability indices to investment decisions. This study is performed for Kish Island in the Persian Gulf.

## I. Introduction

Residential and commercial buildings consume about 32% of the global energy use. They are responsible for about 30% of the total end-use energy-related $CO_2$ emissions, if the indirect upstream emissions are considered [1]. The integration of local distributed energy resources (DERs) and more efficient technologies will result in a further electrification in buildings. From the grid point of view, photovoltaic (PV) systems and electric vehicles (EVs) have an increased grid impact. Furthermore, EVs have a certain flexibility to shift their electricity consumption and they can facilitate the integration of RES. The increase in the penetration of distributed generation and the presence of multiple distributed generators in electrical proximity to one another have brought about the concept of the microgrid [2].

Flexible charging of EVs allows to coordinate the EV charging; hence several coordination scales of the EV integration have been investigated in the literature [3]: The vehicle, building, residential distribution, and transmission grid scale. In [4], a microgrid model has been built with commercial building load, PV array generation, and bi-directional EV charging/discharging stations which aims to minimize the charging cost of EVs and power deviation of the microgrid. An online energy management of PV-assisted charging station under time-of-use pricing is investigated in [5] and a workplace parking lot is simulated as an example to test the method. A new analytical approach to determine the size of PHEV charging stations powered by various levels of commercial PV penetration is proposed in [6] which can find the optimal size of PV-powered charging stations.

Intelligent multi-agent based modeling of power systems is a promising approach to provide a common communication interface for all agents representing the autonomous physical elements in the power system. Furthermore, the distributed nature and potential for modeling autonomous decision making entities in solving complex problems motivates the use of MAS for the operation of the modern power systems through implementing smart grid techniques [7]. According to the aforementioned benefits of using MAS for the operation of microgrids, optimal sizing of the components of microgrids based on the MAS-based energy management of them is necessary. In this regard, in [8], an intelligent multi agent system has been successfully applied to the design of a decentralized energy management system for an autonomous polygeneration microgrid (APM) technology.

In this paper, the optimal sizing problem of an office complex microgrid is considered which is equipped with photovoltaic/fuel cell generation, hydrogen storage, and a PHEV charging station. For this purpose, a multi-agent system is suggested that finds the optimal sizes of the components by mean of particle swarm optimization (PSO) The considered energy management strategy in the proposed MAS controls the charging of PHEVs depending on the hour of the day to reduce cost and avoid overload during peak hours. In this paper, gasoline is considered as the second fuel of the PHEVs.

The rest of this paper is organized as follows: In section II, configuration of the microgrid is presented. Section III describes the multi-agent architecture used for optimal sizing along with the description of components belonging to each agent. Section IV presents the simulation results of the MAS-based architecture used for optimal sizing of the microgrid. Finally, the conclusion of this study is presented in section V.

## II. CONFIGURATION OF THE MICROGRID

Power flow diagram of the proposed microgrid is shown in Fig. 1. According to this diagram, the output power of the photovoltaic unit is equal to $P_{PV}$ that depends on the solar radiation. When the output power of PV arrays is more than electricity demand of the office ($P_{load}$), the surplus power ($P_{PV-sta}$) which can be less than the required charging demand of the station ($P_{sta}$), will be used for charging the PHEVs. After charging the PHEVs, if there is more surplus power, this surplus power ($P_{PV-el}$) will be used in electrolyzer to electrolyze the water and the produced hydrogen which has the power of $P_{el-tank}$, will be stored in the hydrogen tank. If the injected power to the electrolyzer, exceeds electrolyzer's rated power, then the excess energy will circulate in a dump resistor. On the other hand, when the output power of the PV arrays is less than the electricity demand of the office building, then fuel cell consumes the stored hydrogen in the hydrogen tank to the amount of $P_{tank-fc}$ and compensates the shortage of the power production. If the shortage of power exceeds fuel cell's rated power or the stored hydrogen cannot afford that, some fraction of the electricity demand of the office must be shed which leads to loss of load.

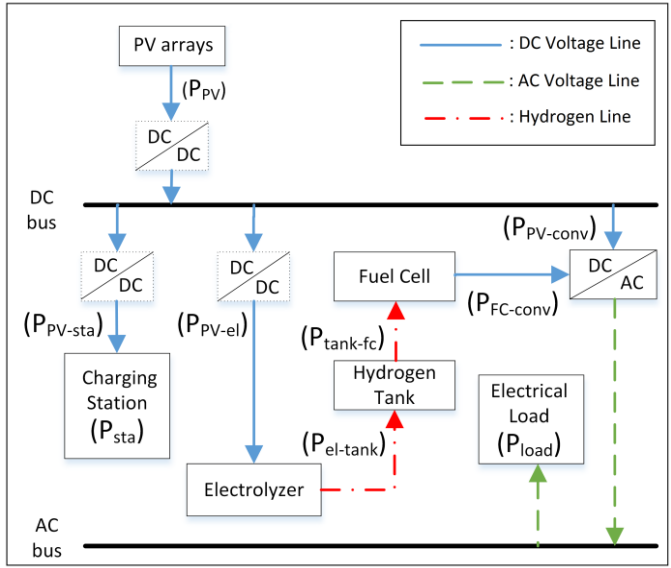

Fig. 1. Schematic diagram of the proposed microgrid system

## III. MULTI-AGENT SYSTEM

The considered MAS for optimal sizing of the proposed microgrid has five agents, namely generation agent, electrical load agent, charging station agent, control agent, and design agent which are organized in three levels. These three levels are presented in Fig. 3.

All the agents associated with the generation or consumption of electricity belong to the field level, which is the lowest level of the considered architecture for the MAS. GA, LA, and SA belong to this level. In the coordination level, coordination of the generation of electricity and consumption of the electrical loads obtains. CA belongs to this level. Design level is the highest level of the proposed architecture that finds the optimal sizes of the components according to the interactions with coordination level such that the electricity demand of the office building and the charging demand of its station are met. In the proposed MAS, DA belongs to this level. The communication channels for data exchange in the proposed MAS is shown in Fig. 2.

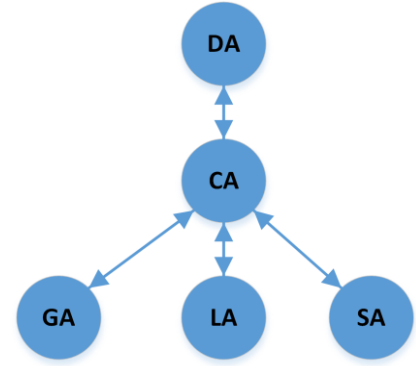

Fig. 2. Data exchange in the MAS

### A. Generation agent

This agent is responsible for management of the resources of the microgrid. Generation resources of the microgrid include PV arrays and the fuel cell. Also, this agent has an electrolyzer and a hydrogen tank to provide hydrogen for the fuel cell and uses a hydrogen tank as a storage unit. Generation agent perceives the amount of power produced by PV arrays and the amount of hydrogen stored in the hydrogen tank by using the sensors. Producing hydrogen by adjusting the operation point of the electrolyzer, generation of electricity in the fuel cell by discharging the hydrogen tank and adjusting the operation point of the fuel cell and also sending the information on the operation of its components to the control agent are its actions.

*1) Components under the supervision of generation agent:* Modeling of the components which are under the supervision of generation agent is presented in the following subsections.

*a) PV arrays:* The output power of the PV generator can be calculated according to the following equation [9]:

$$P_{PV} = \eta_g N_{PV} A_m G_t, \quad (1)$$

where $\eta_g$ is the instantaneous PV generator efficiency that is considered to be 15.4% in this paper [9], $A_m$ is the area of a single module used in the system (m$^2$) which is considered to be 1.9 m$^2$ in this paper [9], $G_t$ is the total global irradiance incident on the titled plane (W/m$^2$), and $N_{PV}$ is the number of modules. In this analysis, each PV array has a rated power of 1 kW. The capital cost of 1 unit is $2000 [9], while the replacement and maintenance costs are taken as $1800, and

$0/yr, respectively. The lifetime of a PV array is taken to be 20 years [10].

*b) Fuel cell:* Proton exchange membrane fuel cell is an environmentally clean power generator which combines hydrogen fuel with oxygen from air to produce electricity. The efficiency of the fuel cell is fed to the computational program as the input. Fuel cell's output power can be defined by the following equation:

$$P_{FC\text{-}conv} = P_{tank\text{-}FC} \times \eta_{FC}. \qquad (2)$$

The capital, replacement, and maintenance costs are taken as $2000 [9], $1500, and $100/yr for a 1 kW system, respectively. The FC's lifetime and efficiency are considered to be 5 years and 50%, respectively [10].

*c) Electrolyzer*: Electrolysis to dissociate water into its separate hydrogen and oxygen constituents has been in use for decades, primarily to meet industrial chemical needs.

The electrolyzer's output power can be calculated by the following equation:

$$P_{el\text{-}tank} = P_{PV\text{-}el} \times \eta_{el}, \qquad (3)$$

In this analysis, a 1 kW system is associated with $1500 capital [9], $1000 replacement, and $15/yr maintenance cost. The electrolyzer's lifetime and efficiency are considered to be 20 years and 75%, respectively [10].

*d) Hydrogen tank:* The energy of hydrogen stored in the tank at time step *t* is obtained by the following equation:

$$E_{tank}(t) = E_{tank}(t\text{-}1) + P_{el\text{-}tank}(t) \times \Delta t - P_{tank\text{-}FC}(t) \times \eta_{storage} \times \Delta t, \qquad (4)$$

where $\eta_{storage}$ is the efficiency of the storage system which is assumed to be 95%, and $\Delta t$ is the duration of each step time which is equal to one hour.

To calculate the mass of stored hydrogen in the tank, the following equation can be used [10]:

$$m_{storage}(t) = E_{tank}(t) / HHV_{H2}, \qquad (5)$$

where $HHV_{H2}$ is the higher heating value of hydrogen which is equal to 39.7 kWh/kg [10]. It is worth mentioning that there are lower and upper limits for amount of the stored hydrogen. It is not possible that the mass of stored hydrogen exceeds the rated capacity of the tank. On the other hand, because of some problems, e.g. hydrogen pressure drop, a small fraction of the hydrogen (here, 5%) may not be extracted. This fraction is the lower limit of the stored hydrogen.

In this analysis, a 1 kg system is associated with $500 capital [9], $450 replacement, and $5/yr maintenance cost. The hydrogen tank's lifetime and efficiency are considered to be 20 years and 95%, respectively [10].

*e) DC/AC converter:* The power electronic circuit used to convert DC into AC is known as inverter. The DC input to the inverter can be from any of the following sources:

- DC output of the PV generation system
- DC output of the fuel cell power system

In this analysis, a 1 kW system is associated with $700 capital [9], $650 replacement, and $7/yr maintenance cost. The converter's lifetime and efficiency are considered to be 15 years and 90%, respectively [10].

*2) Simulation algorithm of the generation agent:* This agent, by coordination of the control agent, interacts with LA and SA in order to supply the electricity and charging demands of the microgrid. In this regard, first CA sends the sizes of PV arrays, electrolyzer, hydrogen tank, fuel cell, and DC/AC converter that are determined by the DA, to the GA and at each time step *t*, asks it for next hour value of the output power of PV arrays. Then, GA sends the requested data to the control agent.

In this paper, it is assumed that GA acts such that when it receives the request of storing the surplus power from the CA, it will use the excess energy in electrolyzer and stores the produced hydrogen in the tank. On the other hand, when the GA receives the request of supplying the shortage of production from the CA, it uses the stored hydrogen in the fuel cell to compensate the shortage of the power production.

## B. Electical load agent

Electrical load agent is a simple reflex agent that aggregates all of the electricity consumptions of the office building except the electricity consumption needed for charging the PHEVs. The electrical loads that LA is responsible for forecasting them are uninterruptable loads and should be supplied subject to a reliability constraint. The aggregated electricity demand of the office building is considered as the only percept of this agent and sending the aggregated electricity demand to the control agent is the only action of it.

## C. PHEV charging station agent

Charging station agent is responsible for charging the PHEVs that arrive at the station on workdays and sends the forecasted charging demands to the control agent. The number of PHEVs that arrive at the station and required charging demand of them are the percepts of it and charging the arrived PHEVs to the desired level of the customer and sending the charging demand and uncharged energy of PHEVs in the predetermined time to the control agent are its actions.

In order to charge a PHEV's battery, electric vehicle supply equipment (EVSE) is considered. For simplicity, it is assumed that PHEVs arrive in time interval $[0,T_1]$ and all PHEVs have the same deadline $T_2$ ($T_2 > T_1$) corresponding to the evening departure time for all employees. Also, it is assumed that each PHEV charges at a constant rate $\psi = 4$ kW until it is charged to the level desired by the customer. The charging level required by each PHEV is uniformly distributed between 0 and 10.4 kW.

In this analysis, SmartCharge-12000 is chosen as EVSE which is fully programmable and open source. This EVSE is designed and manufactured by Electric Motor Werks which is a high-performance electric conversion group [11]. The capital cost of each EVSE is $2000 [11], while the replacement and maintenance costs are taken as $1800, and $20/yr, respectively. The lifetime of each EVSE is taken to be 20 years and the efficiency of each EVSE ($\eta_{sta}$) is considered to be 90% [11].

*1) Simulation algorithm of the station agent:* It is assumed that unmanaged charging profile of the station is available for a year which can be managed to follow the solar power according to the proposed PHEV charging coordination method, thereby reducing the cost of microgrid. This method is such that when the allocated electrical power by CA for the station ($P_{PV\text{-}sta}$) meets the requested charging load ($P_{sta} / \eta_{sta}$), SA does not receive any requests and only sends the information on the operation of its EVSEs to the control agent. On the other hand, When $P_{PV\text{-}sta}$ is less than the requested charging load or it is equal to zero, this agent first checks that whether the current time step is equal to the evening departure time of the PHEVs or not. If this is the case, then this agent calculates the uncharged energy of PHEVs in the predetermined time ($Q_{sta}$) according to (6) and sends it to the control agent.

$$Q_{sta} = (P_{sta} / \eta_{sta}) - P_{PV\text{-}sta}. \qquad (6)$$

On the other hand, if the current time step is not equal to $T_2$, station agent updates the charging demand of the rest of the current day, which are expressed by the set of numbers $\{P_{sta}(t+1), \ldots , P_{sta}(T_2+24n)\}$, by adding a constant value of $((P_{sta} / \eta_{sta}) - P_{PV\text{-}sta}) / ((T_2 + 24n) - t)$ to each of them and sends the updated values of charging load to the control agent at the next time steps. $n$ in the proposed set of numbers represents the number of the day and the term $(T_2 + 24n) - t$ represents the difference of the departure time of a specific day from the current time step.

Furthermore, in both of the above cases, this agent at each time step $t$ sends the information on the operation of its EVSEs to the control agent.

### D. Control agent

Control agent coordinates the interactions between the generation agent and electrical load and station agents. This agent perceives the amount of power produced by PV arrays, charging demands of the office building and PHEV charging station, uncharged energy of PHEVs, and the information on the operation of the microgrid. CA perceives this information by receiving them from GA, LA, and SA. Requesting the GA to store the surplus power produced by PV arrays, Requesting the GA to provide the shortage of power production for supplying the electrical load of the building, allowing the station agent to charge the PHEVs by the surplus power and sending its percepts to the DA are the actions of control agent.

*1) Simulation algorithm of the control agent:* This agent, after receiving a request for operating the microgrid with determined sizes from DA and sending the related sizes to the GA and SA, at each time step $t$, requests GA for the next hour sun radiation, requests LA for the next hour electricity demand of the office building, and requests SA for the next hour charging load needed for charging the PHEVS. After receiving the aforementioned data from GA, LA, and SA, control agent establishes the the following equation:

$$\Delta P = P_{PV} - (P_{load} / \eta_{conv}) - (P_{sta} / \eta_{sta}). \qquad (7)$$

When $\Delta P$ is positive, CA asks GA to store the surplus power of $\Delta P$ and satisfies charging demand of the station. When $\Delta P$ is zero, CA does not send any request to the GA and satisfies charging demand of the station. When $\Delta P$ is negative, in order to check whether output power of the PV arrays can satisfy the electricity demand of the office building or not, this agent establishes the following equation:

$$\Delta P' = P_{PV} - (P_{load} / \eta_{conv}). \qquad (8)$$

In this situation, when $\Delta P'$ is positive, it means that output power of PV arrays can satisfy electrical load of the building and the allocated power for charging PHEVs will be equal to $\Delta P'$. When $\Delta P'$ is equal to zero, it means that output power of PV arrays can satisfy the electrical load of the building, but no power should be allocated for charging the PHEVs. Also, when $\Delta P'$ is negative, it means that output power of PV arrays cannot supply the power needed by LA. Therefore, CA requests the GA to compensate the shortage of power and does not allocate any power to the station. Also, in all of the above cases CA informs the SA about the decision made.

After receiving the information on the operation of the generation and storage components of the microgrid from the GA, when CA finds that electricity demand of the office is not met, it calculates the loss of electrical load of the office building, $Q_{load}(t)$, according to the following equation:

$$Q_{load}(t) = P_{load}(t) - P_{gen}(t), \qquad (9)$$

where $P_{gen}(t)$ is total generation of the considered distributed energy resources.

It is worth mentioning that the aforementioned procedure ensures that the charging demand will be provided only by the surplus power of the PV arrays.

Finally, CA after the receipt of a request for operating the microgrid with determined sizes from DA, by applying the above procedure, sends the information on the operation of the microgrid containing $Q_{load}$ and $Q_{sta}$ to the design agent.

*E. Design agent*

Design agent has the highest level in the architecture of the proposed MAS and it is responsible for optimal sizing of the components of microgrid. The percepts of this agent are same as control agent. Requesting the control agent for sending its percepts at each time step and minimizing the net present cost function of the microgrid through particle swarm optimization with the goal of finding the optimal sizes of the components of the microgrid are the actions of design agent.

*1) System's cost:* In this paper the capital, replacement, and maintenance costs of each component of the microgrid have been considered and net present cost (NPC) is chosen for the calculation of the system's cost. The NPC for a specific device can be expressed by the following equation.

$$NPC_i = N_i \times (CC_i + RC_i \times K_i + MC_i \times PWA(ir, R)), \quad (10)$$

where:

| | |
|---|---|
| $N$ | Number of units and/or the unit capacity (kW or kg) |
| $CC$ | Capital investment cost ($/unit) |
| $K$ | Single payment present worth |
| $RC$ | Replacement cost ($/unit) |
| $MC$ | Maintenance and repair cost ($/unit.yr) |
| $PWA$ | Present worth annual payment |
| $ir$ | Real interest rate (6%) |
| $R$ | Project lifetime (20 years) |

*2) The objective function:* The objective function is sum of all net present costs:

$$NPC = NPC_{PV} + NPC_{el} + NPC_{tank} + NPC_{FC} + NPC_{conv} + NPC_{sta} \quad (11)$$

*3) Reliability:* To evaluate the reliability of supplying the electrical load of the building and charging load of the station, the equivalent loss factor (ELF) index is used. The ELF index for electrical load of the office building and charging load can be calculated by the following equations:

$$ELF_{load} = (1/8760) \sum_{t=1}^{8760} (Q_{load}(t) / P_{load}(t)), \quad (12)$$

$$ELF_{sta} = (1/261) \sum_{t=1}^{261} (Q_{sta}(t) / P_{sta}(t)), \quad (13)$$

where the number 261 indicates the number of evening departure times which is equal to the number of working days in 2016. It is worth mentioning that using the number of evening departure times in (13) is because of that there is no charging load in non-working days and charging load is considered as the deferrable load on working hours.

*4) Simulation algorithm of the design agent:* After the determination of the sizes of the components through PSO by this agent, the determined sizes of the components will be sent to the CA. Accordingly, CA sends the determined sizes of the components to the GA and SA. Then, CA coordinates the GA with LA and SA and sends the information on the operation of the microgrid to the DA. After receiving the aforementioned information, DA calculates the objective function and ELF indices for electricity demand of the office building and charging demand of the station. If $ELF_{load}$ is less than 0.01, $ELF_{sta}$ is less than 0.1, and the energy content of the tank at the end of a year is not less than its initial energy, then design agent selects the obtained sizes as optimum sizes.

## IV. SIMULATION RESULTS

In this section, the proposed MAS-based architecture for optimal sizing of the components of the proposed office complex microgrid is simulated and the optimum combination of the components is calculated. The simulation of the proposed MAS is performed using MATLAB software.

The annual solar radiation data belongs to Kish Island in the Persian Gulf captured by one sample per hour precision that is obtained with HOMER software. For coordinates in HOMER, 26°32' N latitude and 53°58' E longitude are used which are the coordinates of Kish Island.

The annual load curve belongs to a 2000 square meters office building in Kish Island. The annual load pattern of this building has a peak value of 60 kW and is shown in Fig. 3.

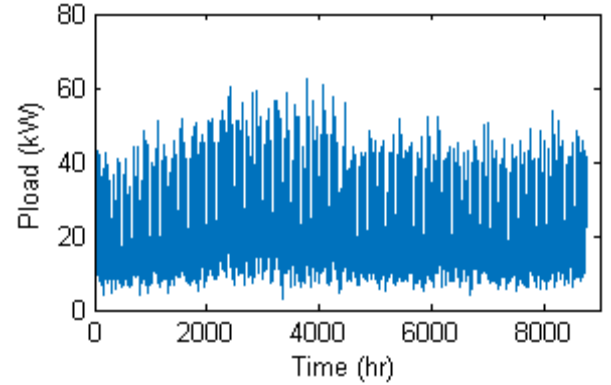

Fig. 3. Annual load curve of the office building

Simulations are done in two scenarios. In scenario1, charging demand of the station is considered as a fixed load and the MAS is simulated without controlling the charging demand. In scenario2, it is assumed that the considered energy management strategy in the proposed MAS controls the charging of PHEVs depending on the hour of the day.

A. Scenario1: In this scenario, charging demand is considered as a fixed load which is shown in Fig. 4. Results of optimal sizing problem through the MAS by considering the charging demand of the station as a fixed load is shown in Table I. In this scenario, the total cost is equal to $1,639,700.

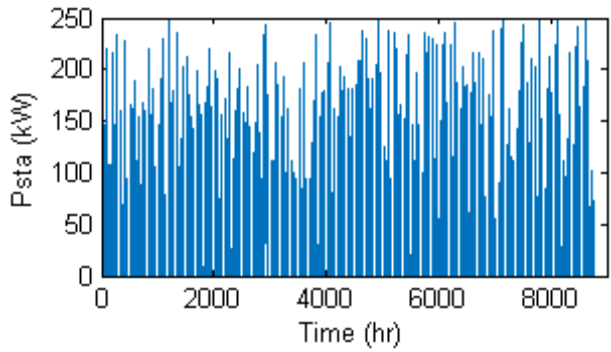

Fig. 4. Annual load curve of the station

TABLE I. OPTIMAL SIZES OF THE COMPONENTS IN SCENARIO1

| PV | Electrolyzer (kW) | Hydrogen tank (kg) | Fuel cell (kW) | DC/AC converter (kW) | EVSE |
|---|---|---|---|---|---|
| 384 | 49.73 | 52.63 | 48.31 | 53.59 | 34 |

*A. Scenario2:* Results of optimal sizing problem through the proposed MAS by considering the charging demand of the station as a deferrable load is shown in Table II. In this scenario, the total cost is equal to $1,571,900.

TABLE II. OPTIMAL SIZES OF THE COMPONENTS IN SCENARIO2

| PV | Electrolyzer (kW) | Hydrogen tank (kg) | Fuel cell (kW) | DC/AC converter (kW) | EVSE |
|---|---|---|---|---|---|
| 318 | 53.48 | 58.39 | 52.21 | 53.97 | 53 |

System's cost in terms of the iterations for the two scenarios is shown in Fig. 5.

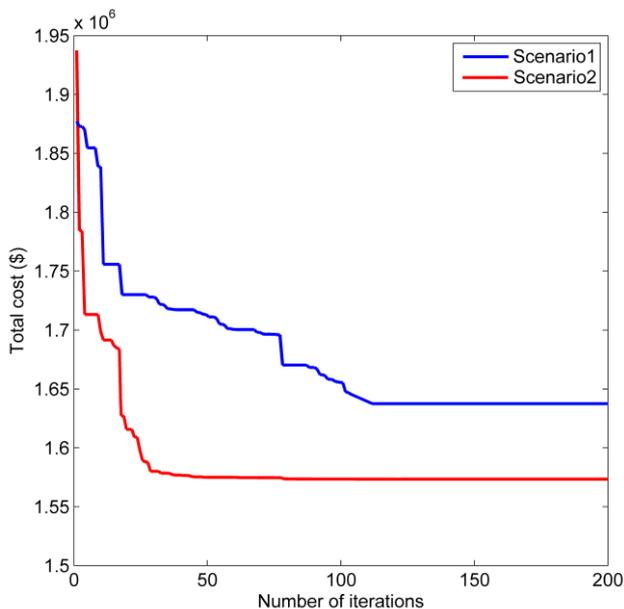

Fig. 5. Total cost in terms of the iterations

Results show that using the proposed flexible charging method in the MAS, decreases the optimal size of PV arrays and increases the optimal size of EVSEs and hydrogen sub-system, which in turn decreases the total cost of the system. This is because of that the cost of reduced number of PVs is more than the cost of added EVSEs and higher capacity hydrogen sub-system. In the proposed MAS, the reason behind decreasing the number of PVs is that it uses an energy management strategy that avoids overload during peak hours, the reason behind increasing the number of EVSEs is that station agent needs more EVSEs so that it can charge the arrived PHEVs in less time, and the reason behind increasing the capacity of hydrogen sub-system is that the number of PV arrays is decreased and at some time steps, more energy should be supplied through hydrogen sub-system to fulfill the power needs of building. Also, there is not much difference in the determined capacity of the converter in the two scenarios, because the electrical load of the building is not deferrable.

V. CONCLUSION

In this paper, development of a multi-agent system for the optimal sizing of an islanded office microgrid is considered. The main problem of renewable energy sources is that they are dependent on environmental conditions, so they cannot cover the demand perfectly. Entering the hydrogen sub-system solves this problem significantly. It is observed that according to the considered energy management strategy in the proposed MAS and controlled charging system, overloading can be avoided by shifting charging demand to off peak hours and therefore, overall cost of the microgrid can be reduced.